\documentclass[conference,10pt]{IEEEtran}

\usepackage{cite}
\usepackage{amsmath,amssymb}
\usepackage{theorem}
\usepackage{rotate}
\usepackage{epsfig,color}
\usepackage{graphicx}
\usepackage{algpseudocode}
\usepackage[ruled]{algorithm}
\usepackage{mdwlist}% Compact list formats \itemize*, \enumerate*
\usepackage{wrapfig}
\usepackage{epstopdf}

\newcommand{\strikeout}[1]{}

\usepackage{tabularx}
\usepackage{gensymb}
\usepackage{epstopdf}
\usepackage{xcolor}
\usepackage{wasysym}
\usepackage{multirow}
\usepackage{tikz}
\usepackage{url}
\usepackage{marvosym}
\usetikzlibrary{arrows,automata}
\usetikzlibrary{shapes,intersections}
\newcommand{\nd}{3cm} % a constant for the node distance
\tikzset{initial text={}}
\tikzset{every picture/.style=semithick} % lines a bit thicker
\tikzset{>=stealth'} %nice arrows
\tikzset{->} %by default, every edge is an arrow
\tikzset{shorten >=1pt}
\definecolor{darkred}{rgb}{0.7, 0.0, 0.0}
\definecolor{darkblue}{rgb}{0.0, 0.0, 0.55}

\newcommand{\highlight}[1]{\textcolor{darkblue}{#1}}
\newcommand{\comment}[1]{ }
\newcommand{\ignore}[1]{ }
\newcommand{\proc}[1]{\textsc{#1}}

\newcommand{\design}{\mathcal{D}}
\newcommand{\shield}{\mathcal{S}}

\newcommand{\gstates}{G}
\newcommand{\fstates}{F}
\newcommand{\ginit}{g_0}

\newcommand{\dinalph}{\Sigma_I}
\newcommand{\dinletter}{{\sigma_I}}

\newcommand{\doutletterprime}{{\sigma_{O'}}}
\newcommand{\dioletter}{{\sigma_{IO}}}

\newcommand{\doutalph}{\Sigma_O}
\newcommand{\dioalph}{\Sigma_{IO}}
\newcommand{\doutalphprime}{\Sigma_{O'}}

\newcommand{\dalph}{\Sigma}
\newcommand{\dletter}{\sigma}

\newcommand{\spec}{\varphi}

\newcommand{\comp}{\circ}

\begin{document}

\title{Shield Synthesis for Real: Enforcing Safety in Cyber-Physical Systems}
\author{
\IEEEauthorblockN{Meng Wu}
\IEEEauthorblockA{\textit{Virginia Tech}\\ Blacksburg, Virginia, USA}
\and
\IEEEauthorblockN{Jingbo Wang, Jyotirmoy Deshmukh, Chao Wang}
\IEEEauthorblockA{\textit{University of Southern California}\\Los Angeles, California, USA}
}

\maketitle

\begin{abstract}
Cyber-physical systems are often safety-critical in that violations of
safety properties may lead to catastrophes. We propose a method to
enforce the safety of systems with real-valued signals by synthesizing
a runtime enforcer called the \emph{shield}.  Whenever the system
violates a property, the shield, composed with the system, makes
correction instantaneously to ensure that no erroneous output is
generated by the combined system.
While techniques for synthesizing Boolean shields are well understood,
they do not handle real-valued signals ubiquitous in cyber-physical
systems, meaning their corrections may be either \emph{unrealizable}
or \emph{inefficient} to compute in the real domain.
We solve the realizability and efficiency problems by analyzing the
compatibility of predicates defined over real-valued signals, and
using the analysis result to constrain a two-player safety game used
to synthesize the shield.  We demonstrate the effectiveness of this
method on a variety of applications, including an automotive
powertrain control system.

\end{abstract}


\section{Introduction}
\label{sec:intro}

A cyber-physical system often needs to continuously respond to
external stimuli with actions under strict timing and safety
requirements; violations of these requirements may lead to
catastrophes.  While formal verification is desirable, in practice, it
can be difficult due to high system complexity, unavailability of
source code, and limited capacity of existing verification tools.
In addition, many systems have started incorporating machine learning
components, which remain challenging to test or
verify~\cite{DreossiFGKRVS19,KatzHIJLLSTWZDK19}.
% 
% Therefore, in the interest of ensuring safety of these systems with
%certainty, runtime enforcement has become an important problem.

Bloem et al.~\cite{BloemKKW15} recently introduced the concept of
\emph{shield}, denoted $\mathcal{S}$, to enforce a specification
$\spec$ of a system $\mathcal{D}$ with certainty.  The goal is to
ensure that the combined $\mathcal{D} \comp \mathcal{S}$ never
violates $\spec$.  If, for example, $\mathcal{D}$ malfunctions and
produces an erroneous output $O$ for input $I$, $\mathcal{S}$ will
correct $O$ into $O'$ \emph{instantaneously} to ensure $\spec(I,O')$
holds even when $\spec(I,O)$ fails.  Here, instantaneously means
correction is made in the same clock cycle.
Furthermore, $\mathcal{S}$ depends solely on $\spec$, which makes it
well-suited for complex $\mathcal{D}$ but small $\spec$, e.g.,
learning-based systems~\cite{WuZWY17,AlshiekhBEKNT18,TuncaliKID18,AvniBCHKP19}.

While the functional specification of $\mathcal{D}$, denoted $\Psi$,
may be large, typically, only a small subset $\spec\subseteq\Psi$ is
\emph{safety-critical}.  Since $\mathcal{S}$ depends on $\spec$, as
opposed to $\Psi$ or $\mathcal{D}$, synthesizing $\mathcal{S}$ from
$\spec$ is more practical than model
checking~\cite{Clarke81,Quiell81}, which decides if $\mathcal{D}$
satisfies $\spec$, or program
synthesis~\cite{Pnueli89,BloemJPPS12,EhlersT14,BloemCGHHJKK14}, which
creates $\mathcal{D}$ from $\Psi$.

Although techniques for synthesizing Boolean shields are well
understood~\cite{BloemKKW15,WuZW16,KonighoferABHKT17}, they do not
work for systems where signals have real values and need to satisfy
constraints such as $x+y\leq 1.53$.  Naively treating the real-valued
constraint as a predicate, or a Boolean variable $P$, may lead to
\emph{loss of information} at the synthesis time and
\emph{unrealizability} at run time.
For example, while the Boolean combination $P \wedge \neg
Q \wedge \neg R$ may be allowed, the corresponding real-valued
constraint may not have solution, e.g., with $P: x+y\leq 1.53$, $Q:
x<1.0$ and $R: y<1.0$.
%
%Therefore, a straightforward combination of the Boolean-level shield
%synthesis techniques with generic constraint solving at the run time
%does not always work.

Even the use of \emph{abstraction refinement} to combine a Boolean
shield with constraint solving does not work.
For example, one may be tempted to block $P\wedge\neg Q\wedge \neg R$
and ask the shield to generate a new solution.  However, since the
shield must be reflexive, i.e., producing $O'$ in the same clock cycle
when the erroneous $O$ occurs, it may be too slow at run time to
recompute a solution.  Even if it is fast enough, the new solution may
still be unrealizable in the real domain.  In general, it is difficult
to bound \emph{a priori} the number of iterations in such an
\emph{abstraction-refinement} loop to meet the strict timing
requirement.

We propose a shield synthesis method to guarantee, with certainty, the
realizability of real-valued signals.
This is accomplished by treating Boolean and real-valued signals
uniformly by adding a set of new constraints.  These constraints take
the form of two automata: a \emph{relaxation automaton}, to capture
the impossible combinations of predicates over signals in $I$ and $O$,
and a \emph{feasibility automaton}, to capture the infeasible
combinations of signals in $O'$.
We use them to restrict the synthesis algorithm formulated as a
two-player safety game, where the \emph{antagonist} controls the
erroneous $O$ and the \emph{protagonist} (shield) controls the
corrected $O'$: the game is won if the protagonist ensures that
$\spec(I,O')$ holds even if $\spec(I,O)$ fails.

\begin{figure}
\centering
\vspace{-2ex}
\includegraphics[width=.95\linewidth]{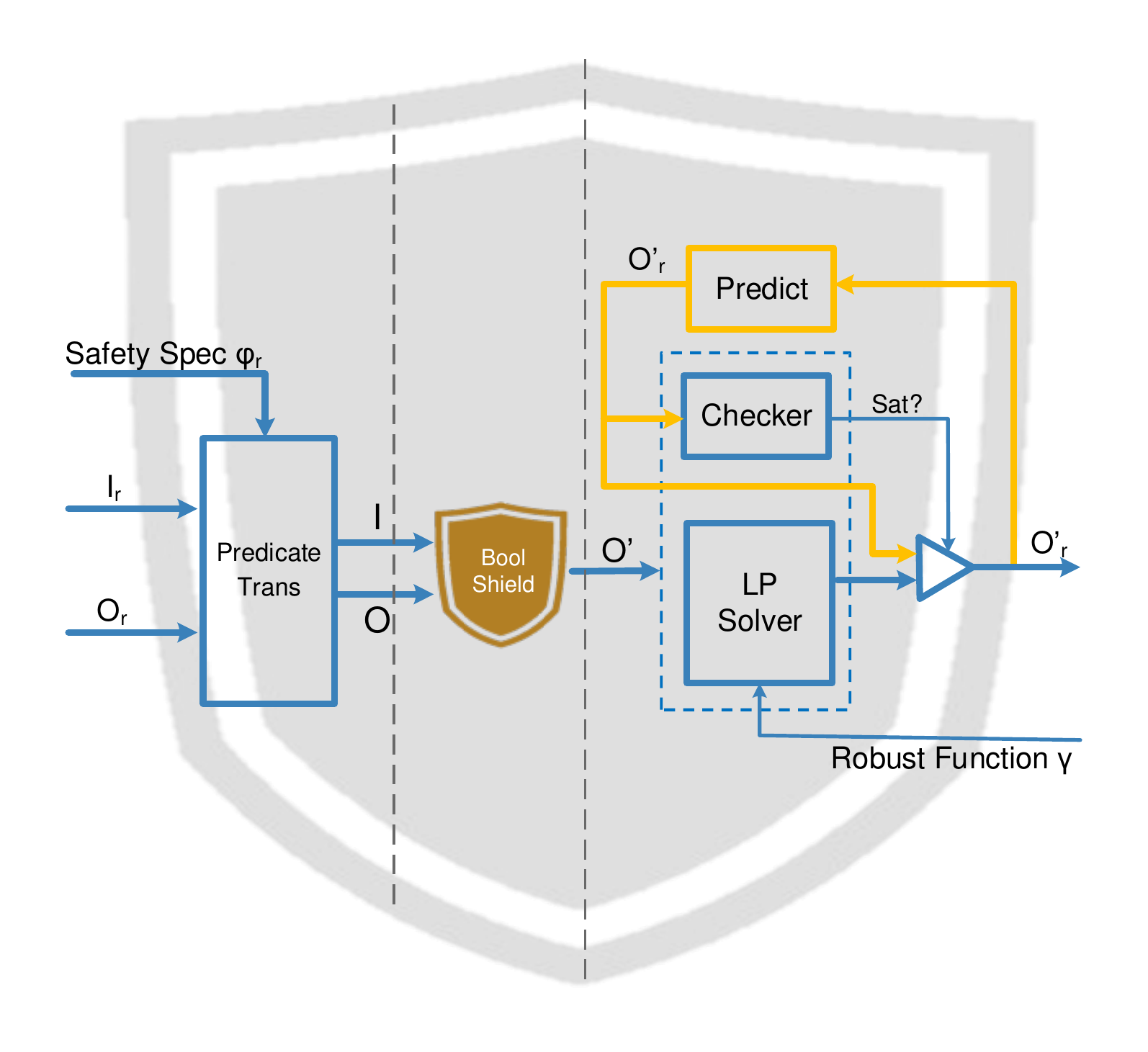}
\vspace{-2ex}

\caption{Overview of the safety shield for real. }
\label{fig:shield_flow}
\end{figure}

The overall flow is shown in Fig.~\ref{fig:shield_flow}, where the
input consists of real-valued $I_r$ and $O_r$ signals and a safety
property $\spec_r$ defined over these signals. 
Internally, the shield $\shield$ has three subcomponents: a converter
from real-valued $I_r/O_r$ signals to Boolean $I/O$ signals, a
converter from Boolean $O'$ signals to real-valued $O'_r$ signals, and
a Boolean shield $\shield(I,O,O')$.
Note that the system, denoted $\mathcal{D}(I_r,O_r)$, is not required
to synthesize the shield: by treating $\mathcal{D}$ as a blackbox, we
ensure that $\mathcal{D}\comp \shield \models \spec_r$ for any
$\mathcal{D}$.

Our shield synthesis algorithm first computes a set $\mathcal{P}$ of
predicates over real-valued signals from $\spec_r$, $I_r$, $O_r$ and
$O'_r$.  Next, it leverages $\mathcal{P}$ to construct the Boolean
abstractions $\spec$, $I$, $O$ and $O'$, as well as the relaxation
automaton $\mathcal{R}(I,O)$ and the feasibility automaton
$\mathcal{F}(O')$.
Using these components, it constructs and solves a safety game where
the antagonist is free to introduce errors to $O$ and the protagonist
must correct them in $O'$.  The winning strategy computed for the
protagonist is the Boolean shield $\shield(I,O,O')$.
At run time, real values are computed for signals in $O'_r$ by solving
a conjunction of constraints based on the Boolean values of signals in
$O$.

To speed up the computation of real values at run time, we also
propose a set of design-time optimizations, which leverage the
information gathered from the shield to simplify the constraints to be
solved at run time.
When there are multiple real-valued solutions, the utility function
$\gamma$ shown in Fig.~\ref{fig:shield_flow}, which defines
a \emph{robustness} criterion, is used to pick the best one.
We also propose a two-phase, \emph{predict-and-validate} technique to
speed up the computation of the real-valued solutions.

We have evaluated the method on a number of applications, including
automotive powertrain control~\cite{JinHSCC2014}, autonomous
driving~\cite{raman2015reactive}, adaptive cruise
control~\cite{nilsson2016correct}, multi-drone fleet
control~\cite{pant2018fly}, generic control~\cite{jin2015mining},
blood glucose control~\cite{roohi2018parameter}, and water tank
control~\cite{AlshiekhBEKNT18}.  Our results show that, in all cases,
the shield can quickly produce real-valued correction signals at run
time.  Furthermore, the use of robustness constraints and two-phase
computation can significantly improve the quality and efficiency of
the real-valued solutions.

To sum up, we make the following contributions:
\begin{itemize}
\item 
We propose a method for synthesizing shields while guaranteeing the
realizability of real-valued signals.
\item
We propose optimizations to speed up the computation and improve the
quality of these correction signals. 
\item 
We demonstrate the effectiveness of the proposed techniques on a
number of applications.
\end{itemize}

The remainder of this paper is organized as follows. First, we review
the basics of shield synthesis in Section~\ref{sec:prelims}.  Then, we
present the technical challenges of extending the Boolean shield to
the real domain in Section~\ref{sec:challenges}.  Details of our 
method for addressing these challenges can be found in
Sections~\ref{sec:bool_shield} and \ref{sec:real_shield}.  Next, we
present our experimental results in Section~\ref{sec:experiment}.  We
review the related work in Section~\ref{sec:related}.  Finally, we
give our conclusions in Section~\ref{sec:conclusion}.

\section{Preliminaries}
\label{sec:prelims}

%\mwnote{more deteails on prior work: construction of safety game, and solving the game}
%
%In this section, we first review the basics of shield synthesis in the
%Boolean domain and then illustrate the high-level idea using an example. 

We assume that the system, $\mathcal{D}$, is a blackbox with input $I$ and
output $O$.  When $\mathcal{D}$ malfunctions, it produces some
erroneous values in $O$.  The shield, $\shield$, takes both $I$ and
$O$ as input and returns $O'$ as output.  Whenever
$\mathcal{D}\models\spec$, the shield returns $O'=O$; and when
$\mathcal{D}\not\models\spec$, the shield needs to compute correction
$O'$ for $O$.  Following Bloem et al.~\cite{BloemKKW15}, we treat the
correction computation as a two-player safety game.

\vspace{1ex}
\noindent
\textbf{Safety Game}~~
%
%\subsection{Safety Game}
%
The antagonist controls the alphabet $\dioalph$ and the protagonist
controls the alphabet $\doutalphprime$.
The game is a tuple $\mathcal{G} = (G,
g_0, \dioalph \times \doutalphprime, \delta_\mathcal{G}, F)$, where
$G$ is a set of states, $g_0$ is the initial state,
$\dioalph\times \doutalphprime$ is the combined alphabet,
$\delta_\mathcal{G}:
G \times \dioalph \times \doutalphprime \rightarrow G$ is the
transition function, and $F$ is the set of unsafe states.
In each state $g \in G$, the antagonist chooses a letter
$\dioletter \in \dioalph$ and then the protagonist chooses a letter
$\doutletterprime \in \doutalphprime$, thus leading to state $g'
= \delta_\mathcal{G} (g, \dioletter, \doutletterprime)$.
The resulting state sequence $g_0g_1 ...$ is called a play. A play is
\emph{winning} for the protagonist when it visits only the safe states.

The game may be solved using the classic algorithm of
Mazala~\cite{Mazala01}, which computes ``attractors'' for a subset of
safe states $(G\setminus F)$ and unsafe states $F$.
A winning \textit{region} $\mathcal{W}$ is a subset of $(G\setminus
F)$ states within which the protagonist has a strategy to win.
A winning \textit{strategy} is a function $\omega:
G \times \dioalph \rightarrow \doutalphprime$ that ensures the
protagonist always wins.   
The shield $\shield$ is an implementation of the winning strategy.

\vspace{1ex}
\noindent
\textbf{Boolean Shield}~~
%
%\subsection{Boolean Shield}
%
It is a tuple $\mathcal{S} = (S,
s_0, \dioalph, \doutalphprime, \delta_\mathcal{S}, \lambda_{\mathcal{S}})$,
where $S$ is a set of states, $s_0$ is the initial state,
$\delta_{\mathcal{S}}: S \times \dioalph \rightarrow S$ is the
transition function, and $\lambda_{\mathcal{S}} (S, \dioletter)
= \doutletterprime$ is the output function. Here,
$\lambda_{\mathcal{S}}$ implements the winning strategy $\omega$ in
the game $\mathcal{G}$.
Assume the system is $\mathcal{D} = (Q,
q_0, \dinalph, \doutalph, \delta_{\mathcal{D}}, \lambda_{\mathcal{D}}
)$, where $Q$ is the set of system states, $q_0$ is the initial state,
$\delta_{\mathcal{D}}: Q \times \dinalph \rightarrow Q$ is the
transition function, and $\lambda_{\mathcal{D}}:
Q \times \dinalph \rightarrow \doutalph$ is the output function.
The composition $\design \comp \shield$ is $( QS,
qs_0, \dinalph, \doutalphprime, \delta_{\mathcal{D \comp
S}}, \lambda_{\mathcal{D \comp S}} )$, where $QS = Q\times S$, the
initial state is $qs_0 = (q_0,s_0)$, the transition function is
$\delta_{\mathcal{D\comp S}}: QS \times \dinalph \rightarrow QS$, and
the output function is $\lambda_{\mathcal{D\comp S}}$.

Given a state $qs=(q,s)$, the next state $qs' = (q',s')$ is computed
by $\delta_{\mathcal{D\comp S}}(qs,\dinletter)$ as follows: $q'
= \delta_{\mathcal{D}}(q,\dinletter)$ and $s'
= \delta_{\mathcal{S}}(s,\dinletter \comp \lambda_{\mathcal{D}}(\dinletter)
)$,
and $\doutletterprime$ is computed by $\lambda_{\mathcal{D\comp
S}}(qs,\dinletter)$ as follows:
$\lambda_{\mathcal{S}}(s, \dinletter \comp \lambda_{\mathcal{D}}(\dinletter)
)$.

%\vspace{-3ex}
%\subsubsection{Correctness and Minimum Deviation}

%Let $\mathcal{L}(\design \comp \shield)$ be the language (set of
%input-output sequences) generated by the composed system and let
%$\mathcal{L}(\spec)$ be the language accepted by the 
%specification $\spec$.  We explain why $\mathcal{L}(D \comp
%S) \subseteq \mathcal{L} (\spec)$.
%%
%First, when $\mathcal{D}(I,O) \models \spec(I,O)$, the shield returns
%$\doutletter' = \doutletter$ and thus ensures $\mathcal{D\comp
%S} \models \spec(I,O')$.
%%
%Second, when $\mathcal{D}(I,O) \not\models \spec(I,O)$, the shield
%generates $O'$ by making corrections to $O$ so as to ensure that
%$\spec(I,O')$ holds.
%%
%Thus, in both cases, $\mathcal{D\comp S} \models \spec(I,O')$.

If there are multiple ways of changing $O$ to $O'$ to satisfy
$\spec(I,O')$, the shield must choose the one with minimum difference
between $O$ and $O'$.  The difference may be measured in Hamming
Distance~\cite{BloemKKW15}: when $\mathcal{D}\models \spec$,
$\mathit{HD}(O,O')=0$; and when $\mathcal{D}\not\models\spec$,
$\mathit{HD}(O,O')$ is minimized.

\vspace{1ex}
\noindent
\textbf{Example}~~
%
%\subsection{Example}
\label{sec:example}
Consider the following two formulas in LTL~\cite{Pnueli77}: 
%$\mathsf{G} \Big( A \Rightarrow B_1 \Big)$ and
%$\mathsf{G} \Big( A \wedge \mathsf{X}(\neg A) \Rightarrow \mathsf{G} B_2 \Big)$,

{\small
\vspace{-1ex}
\begin{align}
\mathsf{G} \Big( A \Rightarrow B_1 \Big) \nonumber \\
\mathsf{G} \Big( A \wedge \mathsf{X}(\neg A) \Rightarrow  B_2 \mathsf{U} A\Big) \nonumber
\end{align}
\vspace{-2ex}
}

\noindent
where $\mathsf{G}$ means \emph{Globally}, $\mathsf{X}$
means \emph{Next}, $\mathsf{U}$ means \emph{Until}, Boolean variable $A$ is an input signal, while $B_1$ and $B_2$ are output signals of $\mathcal{D}$.
Fig.~\ref{fig:specification} shows the corresponding automaton representation,
% of
%these two formulas in LTL~\cite{Pnueli77}, 
where 0 is the initial state and 2 is an unsafe state. Note that the
transition labels are on the edges.

Fig.~\ref{fig:boolean-shield} shows the shield generated by existing
techniques~\cite{BloemKKW15,WuZW16,KonighoferABHKT17}, which takes
signals $A$, $B_1$ and $B_2$ as input, and return the modified signals
$B_1'$ and $B_2'$ as output.  Here, labels are on nodes instead of
edges: they are conditions under which transitions go to destination
nodes. Furthermore, $B_1=B'_1$ means the two signals have the same
value.

The shield ensures that $A$, the input signal of $\mathcal{D}$, and
$B'_1,B'_2$, the modified output signals of $\mathcal{D}$, always
satisfy the specification in Fig.~\ref{fig:specification}.  At the
same time, the deviation between $B_1,B_2$ and $B'_1,B_2'$ is
minimized.

\begin{figure}
\hspace{-4ex}
\centering
\scalebox{0.65}{\begin{tikzpicture}[auto,node distance=\nd]
\node[state, initial=left]         at  (5,0)     (S0) {0};
\node[state]         at  (8,0)       (S1) {1};
\node[state,accepting]  at  (6.5,-2)  (S2) {2};

\path

 (S0)  edge [loop above]  
 node [xshift=-10mm, yshift=-5mm]  
 {$A \wedge B_1$} (S0) 

 (S0)  edge [bend left =15]
 node [xshift=1mm]  
 {$\neg A$} (S1) 

 (S1)  edge [bend left =15]
 node [xshift=1mm]  
 {$A \wedge B_1$} (S0) 

 (S0)  edge []
 node [xshift=-14mm, yshift=-5mm]  
 {$A \wedge \neg B_1$} (S2) 

(S1)  edge [red,thick,dashed]
 node [xshift=0mm, , yshift=2mm, align=left]  
 {${\color{red}\neg A \wedge \neg B_2}$ \\ ${\color{black}\vee A \wedge \neg B_1}$} (S2) 

 (S2)  edge [loop below]  
 node [xshift=10mm, yshift=5mm]  
 {$True$} (S2)

  (S1)  edge [loop above]  
 node [xshift=10mm, yshift=-5mm]    
 {$\neg A \wedge B_2$} (S1)
;
\end{tikzpicture}}
\caption{Boolean safety specification.}
\label{fig:specification}
\end{figure}
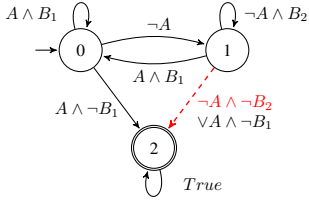
\begin{figure}
\centering
\scalebox{0.85}{\begin{tikzpicture}[auto,node distance=\nd]
\node[state, initial=left]  at  (5,0)     (S0) {0};
\node[state]                at  (8,0)     (S1) {1};
\node[state]                at  (12,0)    (S2) {2};
\node[state]                at  (6.5,-2)  (S3) {3};

\path
 (S0)  edge [loop above]  
 node [ align=right] 
 {\scriptsize $A \wedge B_1 \wedge B_1'$ \\ \scriptsize $\wedge (B_2 = B_2')$} (S0) 

 (S0)  edge [bend left =15]
 node [xshift=1mm]  
 {} (S1) 

 (S1)  edge [bend left =15]
 node [xshift=1mm]  
 {} (S0) 

 (S1)  edge [bend left =15, red,dashed]
 node [xshift=-3mm, yshift=-1mm, align=right] 
 {} (S2) 

 (S2)  edge [bend left =15]
 node [xshift=1mm]  
 {} (S1)

 (S2)  edge [loop right]  
 node [xshift=-16mm, yshift=-10mm, align=left]  
 {\scriptsize $\neg A \wedge (B_1 = B_1')$ \\ \scriptsize $\wedge \neg B_2 \wedge B_2'$} (S2)

 (S1)  edge [bend left =15]
 node [xshift=1mm]  
 {} (S3) 

(S3)  edge [bend left =15]
 node [xshift=1mm]  
 {} (S1)

(S0)  edge [bend left =15]
 node [xshift=1mm]  
 {} (S3) 

(S3)  edge [bend left =15]
 node [xshift=1mm]  
 {} (S0) 

 (S2)  edge [bend left =15]
 node [xshift=1mm]  
 {} (S3)

 (S3)  edge [loop below]  
 node [xshift=12mm, yshift=7mm, align=left] 
 {\scriptsize $A \wedge \neg B_1 \wedge B_1'$ \\ \scriptsize $\wedge (B_2 = B_2')$} (S3)

  (S1)  edge [loop above]  
 node [xshift=13mm, yshift=-6mm, align=left] 
 {\scriptsize $\neg A \wedge (B_1 = B_1') $ \\ \scriptsize $\wedge B_2 \wedge B_2'$} (S1)

 (S2)  edge [bend right =45]
 node [xshift=1mm]  
 {} (S0)
;
\end{tikzpicture}}
\caption{Boolean shield for Properties in Fig.~\ref{fig:specification}.} 
\label{fig:boolean-shield}
\end{figure}
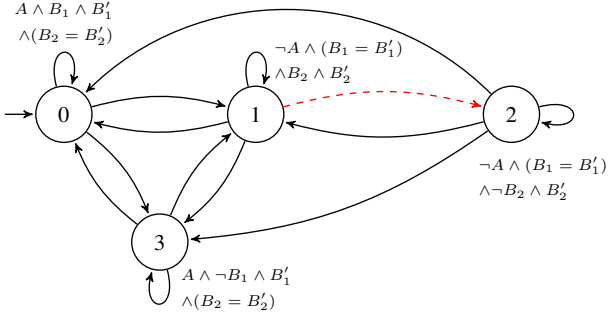

The red dashed edge in Fig.~\ref{fig:boolean-shield} illustrates a
scenario where $\mathcal{D}$ violates the specification by setting
$B_2$ to \textit{false} while moving from a state where $A$ is true to
a state where $A$ is false, as represented by the red dashed edge in
Fig.~\ref{fig:specification}.  The shield, upon detecting this
violation, responds instantaneously by setting $B'_2$ to
\textit{true}.  It allows the specification to be satisfied by the
modified output ($B'_2$).
As for $B'_1$, whose value does not matter, the shield maintains $B'_1
= B_1$ to minimize the deviation.

\section{Technical Challenges}
\label{sec:challenges}

Using a Boolean shield to generate real-valued correction signals has
two problems: realizability of the Boolean predicates, and quality of
the real-valued signals.

\subsection{Realizability of the Boolean Predicates}

The Boolean specification in Section~\ref{sec:example} are
abstractions of the real-valued LTL properties below, which in turn
are abstractions of properties of an automotive powertrain control
system~\cite{JinHSCC2014} expressed in Signal Temporal Logic
(STL~\cite{MalerN04}).

\vspace{-1ex}
{\small
\begin{align}
\mathsf{G} \big( l\!=\!\textsf{power} \Rightarrow |\mu| < 0.2 \big) \nonumber \\
\mathsf{G}  \Big( l\!=\!\textsf{power} \wedge \mathsf{X}(l\!=\!\textsf{normal}) \Rightarrow  \big( |\mu| < 0.02 \big)  \mathsf{U}  \big(l\!=\!\textsf{power}\big) \Big)\nonumber
\end{align}
}

%\vspace{-2ex}
%{%\footnotesize
%\begin{tabular}{lp{.9\linewidth}lp{.1\linewidth}}
%  $A$:   & System is in Power mode $l\!=\!\textsf{power} $\\
%  $B_1$: & the normalized error of air-fuel ration $|\mu| < 0.2$ \\
%  $B_2$: & the normalized error of air-fuel ration $|\mu| < 0.02$   
%\end{tabular}
%}
%\vspace{1ex}

The input signal $l$ denotes the system mode, which may be
\textsf{normal} or \textsf{power}. The output signal $\mu$ is the
normalized error of the air-fuel (A/F) ratio inside an internal
combustion engine.  Let $\lambda$ be the A/F ratio and $\lambda_{ref}$
be a reference value, then $\mu = (\lambda - \lambda_{ref}) /
\lambda_{ref}$.  Since $\mu$ affects other parts of the systems, it
must be kept in certain regions depending on the system mode.

The first property says that $|\mu|$ should stay below 0.2 in
the \textsf{power} mode.  
The second property says that, after the system changes from the
\textsf{power} mode to the \textsf{normal} mode, $|\mu|$ should 
stay below 0.02.
In the Boolean versions, $A$ denotes whether the system is in
the \textsf{power} mode, while $B_1$ and $B_2$ denote $|\mu|<0.2$ and
$|\mu|<0.02$, respectively.  The combination $\neg B_1 \wedge B_2$ is
unrealizable, because $|\mu|$ cannot be both greater than $0.2$ and less
than $0.02$.

However, the shield synthesized by existing methods is not aware of this
problem, and thus may produce combinations of Boolean values that are
not realizable in the real domain. 
As shown by the red edge in Fig.~\ref{fig:boolean-shield}: if the
shield's input is $\neg A \wedge \neg B_1 \wedge \neg B_2$, the
shield's output will be $\neg B_1' \wedge B_2'$, despite that $|\mu'|
\geq 0.2 \wedge |\mu'| < 0.02$ is unsatisfiable.

%Although one may be tempted to use \emph{iteration}, e.g., blocking
%the bad solution and re-computing a solution, in general, there is
%no guarantee that the new solution is realizable either.  Since there
%is no way of bounding the number of iterations \emph{a priori}, the
%response time of the shield will no longer be bounded at run time.
%This contradicts to the strict timing requirements of
%systems where correction is expected to occur
%instantaneously.
%
%Furthermore, it is also possible for the shield (which implements only
%one of the possible winning strategies) to run out of solutions, even
%though a realizable solution may be found by another shield.  However,
%re-synthesizing the shield at run time would be even more
%time-consuming.

We solve this problem by checking the compatibility of the predicates
at the synthesis time, to guarantee their realizability at run time.
Details will be presented in Section~\ref{sec:bool_shield}.

\subsection{Quality of the Real-valued Output}

Even if the Boolean values are realizable, the real-valued solution
computed by a generic solver may not be of high quality.  Assume that
all predicates are linear arithmetic constraints, the output of a
Boolean shield would be a conjunction of constraints.  As shown in
Fig.~\ref{fig:shield_flow}, the back-end may convert $O'$, the Boolean
output, to $O'_r$, the real-valued output, by solving a linear
programming (LP) problem.

\begin{figure}
	\centering
	\includegraphics[width=\linewidth]{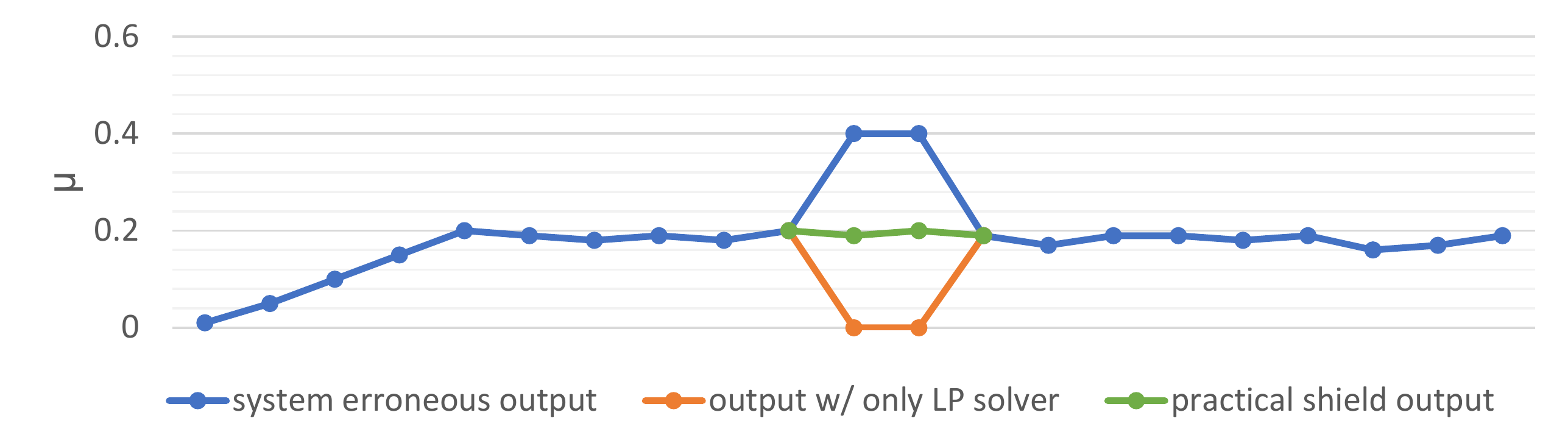}
	\caption{Importance of the smoothness in real-valued correction signals.}
	\label{fig:real_ex}
\end{figure}

%However, it may not produce a reasonable output. 
Consider $\mathsf{G} ( A \Rightarrow B )$, which abstracts $\mathsf{G}
\big( l\!=\!\textsf{power} \Rightarrow |\mu| < 0.2 \big)$.
Suppose the original system's output violates the property $|\mu| <
0.2$ as shown by the blue line in Fig.~\ref{fig:real_ex}, where the
two erroneous values are in the middle.  The correction computed by an
LP solver may be any of the infinitely many values in the interval
(-0.2, +0.2), including -0.19 and 0. However, neither of these two
values may be acceptable in a real system, which expects the signal
to be \emph{stable}, not \emph{arbitrary}.
%\mwnote{want to emphasis LP solver will give a const, which is not practical.}

Ideally, we want to generate real-valued signals that are smooth, and
consistent with physical laws of the environment, e.g., the green line
in Fig.~\ref{fig:real_ex}.
Toward this end, we leverage a utility function, $\gamma$, to impose
\emph{robustness}  in addition to \emph{correctness}  constraints.  With
both types of constraints, the LP solver can generate values of high
quality.

We also propose a technique to speed up the computation of real
values. The intuition is that system dynamics may be approximated
using (linear) regression, which predicts the current
value of a signal based on its values in the recent past.  Thus, we
develop a fast runtime prediction unit to guess the value, followed by a fast
validation unit to check its validity.  If the predicted value is
valid, it will serve as the shield's output.  Otherwise, we invoke the
LP solver.  Details will be presented in
Section~\ref{sec:real_shield}.

\section{Synthesizing the Boolean Shield}
\label{sec:bool_shield}

In this section, we present our method for ensuring the realizability
of the shield's output signals.
%Unlike the \emph{iterative blocking} approach mentioned in the
%example from last section, our method analyzes the realizability
%problem during the shield synthesis. Thus, it guarantees that real
%values can always be computed at run time.
The idea is to check the compatibility of predicates inside the
game-based algorithm for synthesizing the Boolean shield.  To improve
efficiency, we check predicate combinations only when they are
involved in compute the winning strategy.

\begin{algorithm}[t]
{\footnotesize
\caption{\footnotesize Synthesizing a realizable Boolean shield $\shield_{bool}$ from $\spec_{r}$.}
\label{alg:synth}
\begin{algorithmic}[1]
\State \highlight{Let $\mathcal{P}$ be the set of predicates over real-valued variables in $\spec_{r}$;}
\State \highlight{Let $\spec$, $I$, $O$, $O'$ be Boolean abstractions of $\spec_{r}$, $I_{r}$, $O_{r}$, $O'_{r}$ via $\mathcal{P}$;}

\Function{SynthesizeBool~}{ $\mathcal{P}$, $I$, $O$, $O'$ }
\State $\mathcal{Q}(I,O') \leftarrow \proc{genCorrectnessMonitor}( \spec )$
\State $\mathcal{E}(I,O,O') \leftarrow \proc{genErrorAvoidingMonitor}( \spec )$
\State $\mathcal{G}\leftarrow \mathcal{Q} \comp \mathcal{E}$   
\State $\mathcal{W} \leftarrow \proc{computeWinningStrategy}(\mathcal{G})$

\State \highlight{$\mathcal{R}(I,O) \leftarrow \proc{genRelaxationAutomaton}(P,I,O,\mathcal{W})$}
\State \highlight{$\mathcal{F}(O')  \leftarrow \proc{genFeasibilityAutomaton}(\mathcal{R})$}

\State \highlight{$\mathcal{G}_\mathit{r} \leftarrow \mathcal{W} \comp \mathcal{R} \comp \mathcal{F}$}
\State \highlight{$\omega_\mathit{r} \leftarrow \proc{computeWinningStrategy}(\mathcal{G}_\mathit{r})$}

\State {$\shield_{bool}(I, O, O') \leftarrow \proc{implementShield}(\omega_\mathit{r})$}

\State {\bf return} $\shield_{bool}$
\EndFunction
\end{algorithmic}
}
\end{algorithm}

Algorithm~\ref{alg:synth} shows the procedure, where blue highlighted
lines address the \emph{realizability} issue while the remainder
follows the classic algorithm in the prior
work~\cite{BloemKKW15,WuZW16,KonighoferABHKT17}.
First, it creates $\mathcal{P}$, the set of predicates from the
real-valued specification $\spec_{r}$.  Then, it uses $\mathcal{P}$ to
compute a Boolean abstraction of $\spec_{r}$, denoted $\spec$.  Next,
it uses $\spec$ to formulate a two-player safety game $\mathcal{G}$
where the antagonist controls $I$ and $O$, the protagonist controls
$O'$, and $\mathcal{W}$ is the winning region where the protagonist
may win the game.

Since the construction of the safety game $\mathcal{G}$ is part of the
prior work, we refer to Bloem et al.~\cite{BloemKKW15} and Wu et
al.~\cite{WuZW16} for details.  Here, it suffices to say that
$\mathcal{G}$ is a synchronous composition of $\mathcal{E}$, an
error-avoiding monitor that outlines all possible ways in which the
antagonist may introduce errors in $O$ and the protagonist may
introduce corrections in $O'$, and $\mathcal{Q}$, a correctness
monitor that ensures $\spec(I,O')$ always holds.

Since a winning strategy in $\mathcal{W}$ may not be realizable in the
real domain, our next step is to compute a strategy $\omega_r$ based
on $\mathcal{W}$ while ensuring correction signals produced by
$\omega_r$ are always realizable.  Toward this end, we introduce two
additional automata: the \emph{feasibility} automaton
$\mathcal{F}(O')$ and the \emph{relaxation} automaton
$\mathcal{R}(I,O)$.
Specifically, $\mathcal{F}$ is used to identify and remove the
infeasible edges in $\omega$, i.e., corrections in $O'$ with no
real-valued solutions.  $\mathcal{R}$ is used to identify and remove
the unrealistic errors in $I$ and $O$, i.e., errors that are
impossible and thus will not occur in the first place.

In other words, $\mathcal{F}$ restricts the search to realizable
solutions, and $\mathcal{R}$ allows us to have more freedom while
computing the winning strategy.  Thus, the new game
$\mathcal{G}_\mathit{r}$ is a composition of $\mathcal{W}$,
$\mathcal{R}$ and $\mathcal{F}$.  Based on the winning strategy
$\omega_r$ computed from $\mathcal{G}_r$, we can construct a shield
$\shield_{bool}$ that is guaranteed to be realizable at run time.

In the remainder of this section, we illustrate the details while
focusing on the highlighted lines in Algorithm~\ref{alg:synth}.

\subsection{Computing the Predicates}
\label{sec:predicates}

%\mwnote{Define predicate $\mathcal{P}$ first, which should be a subset of $I \cup O$, and $\dalph_p \in \dalph$. $\gamma^p$ is the STL formula of $\dalph_p$. $\gamma^p$ is unsatisfiable if there does not exist $I_r$ and $O_r$...}

$\mathcal{P}$ is the set of predicates over real-valued signals used
in $\spec_r$, where $\spec_r$ is expressed in Signal Temporal Logic
(STL).  In addition to the LTL operators, STL also has dense time
intervals associated with temporal operators and constraints over
real-valued variables.

Consider the STL formulas below, which come from the powertrain
control system~\cite{JinHSCC2014} without modification.

\vspace{-1ex}
{\footnotesize
\begin{align}
\mathsf{G}_{[\tau_s,T]} \big( l\!=\!\textsf{power} \Rightarrow |\mu| < 0.2 \big) \nonumber \\
\mathsf{G}_{[\tau_s,T]}  \Big( l\!=\!\textsf{power} \wedge \mathsf{X}(l\!=\!\textsf{normal}) \Rightarrow  \mathsf{G}_{[\eta,\frac{\varsigma}{2}]} \big( |\mu| < 0.02 \big) \Big) \nonumber 
\end{align}
}

\vspace{-1ex}
\noindent
Here, $\mathsf{G}_{[\tau_1,\tau_2]}$ is the temporal operator
augmented with time interval $[\tau_1,\tau_2]$, $l$ is the system
mode, and $\mu$ is the normalized error of the air-fuel ratio.
The first property says that $|\mu|$ should stay below 0.2 immediately
after the system switch to the \textsf{power} mode, i.e., between time
$\tau_s$ and time $T$.
The second property says that, when it switches from the
\textsf{power} mode  to the \textsf{normal} mode, $|\mu|$ should
settle down to below 0.02 after time $\eta$ and before time
$\frac{\varsigma}{2}$.

To compute $\mathcal{P}$, first, we convert each time interval to a
 conjunction of linear constraints, e.g., by using a time variable $t$
 to represent the bounds in intervals $[\tau_s, T]$ and
 $[\eta, \frac{\varsigma}{2}]$.

\vspace{1ex}
{%\footnotesize
\begin{tabular}{lp{.3\linewidth}lp{.4\linewidth}}
  $T_1$:&   $(t \geq \tau_s)$       &   $T_2$:&$   (t \leq T)$ \\
  $T_3$:&   $(t \geq \eta)$         &   $T_4$:&$   (t \leq \frac{\varsigma}{2})$\\
\end{tabular}
}
\vspace{1ex}

\noindent
Next, we convert the constraints over real-valued variables to
predicates.  From the running example, we will produce the following
predicates:

\vspace{1ex}
{%\footnotesize
\begin{tabular}{lp{.3\linewidth}lp{.4\linewidth}}
  $L_1$:&   $(l = \mathsf{power})$  &   $L_2$:&$   (l = \mathsf{normal})$ \\
  $M_1$:&   $(|\mu| < 0.2)$         &   $M_2$:&$   (|\mu| < 0.02)$\\
\end{tabular}
}

\subsection{Computing the Boolean Abstractions}

After the set $\mathcal{P}$ of predicates is computed, we use it to
compute the Boolean abstractions of $\spec_{r}$, $I_{r}$, $O_{r}$ and
$O'_{r}$.  This step is straightforward.  To compute $\spec$ from
$\spec_r$, we traverse the abstract syntax tree (AST) of $\spec_r$
and, for each AST node $n$ that corresponds to a real-valued predicate
$P\in \mathcal{P}$, we replace $P$ with a new Boolean variable $v_P$.

To compute $I$ from $I_r$, we traverse the predicates in $\mathcal{P}$
and, for each predicate $Q\in \mathcal{P}$ defined over some
real-valued signals in $I_r$, we add a new Boolean variable $v_Q$ to
$I$. Similarly, $O$ and $O'$ are also computed from $O_r$ and $O_r'$ by 
creating new Boolean variables.

\subsection{Computing the Relaxation Automaton}
%\mwnote{need to include definition of following literals in section~\ref{sec:prelims}.}

The relaxation automaton $\mathcal{R}$ aims to identify impossible
combinations of $I$ and $O$ values, and since they will never occur in
the shield's input, there is no need to make corrections in the
shield's output.  There may be two reasons why a value combination is
impossible:
\begin{enumerate}
\item
The values of real-valued predicates are incompatible, e.g., as in
$|\mu|<0.02$ and $|\mu|>0.2$.
\item
The values are not consistent with physical laws of the environment,
e.g., time never travels backward. For example, with respect to the
time interval $[\tau_s,T]$, the transition from $T_1 \wedge T_2$ to
$\neg T_1 \wedge T_2$ is impossible.
\end{enumerate}
In addition, our method allows users to provide more constraints to
characterize physical laws of the environment or their understanding
of the behaviors of the system $\mathcal{D}$.

States in the relaxation automaton $\mathcal{R}$ are divided into two
types: \emph{normal} states and \emph{impossible}
states. Here, \emph{normal} means the $I/O$ behavior of the system
$\mathcal{D}$ may occur, whereas \emph{impossible} means it will never
occur.  Since impossible $I/O$ behavior will never occur in the
shield's input, the shield may treat it as \emph{don't-care} and thus
have more freedom to compute the winning strategy.

\vspace{1ex}
\noindent
\textbf{Example}~~
Fig.~\ref{fig:relax} shows the relaxation automaton for our running
example.  Here, the dashed edges come from the physical laws (time
never travels backward), while the solid edges comes from the
compatibility of real-valued predicates defined over $l$ and $\mu$.
In particular, the combination $\neg M_1 \wedge M2$ is identified as
impossible, because $|\mu|$ cannot be greater than 0.2 and less than
0.02 at the same time.

\begin{figure}
	\centering
	\begin{minipage}[b]{\linewidth}
		\centering
		\scalebox{0.65}{\begin{tikzpicture}[auto,node distance=\nd]
\node[state,initial] at  (0,0)       (S0) {};
\node[dashed, state]         at  (3.5,0)     (S2) {};
\node[dashed, state]         at  (1.75,1.5)     (S3) {};
\node[state,accepting]         at  (1.75,-1.5)     (S1) {impossible};
\path
(S0) edge [loop above] 
     node [xshift=-25mm,yshift=-5mm,align=right] 
     {($\neg T_1 \wedge T_2$) $\vee$ ($M_1 \vee \neg M_2 $)} (S0)

(S0) edge [bend right = 20]             
     node [xshift=-45mm,yshift=-5mm, align=center] 
     {($\neg M_1 \wedge M_2$) $\vee$ ($\neg T_1 \wedge \neg T_2$)} (S1)

(S0) edge [dashed, bend left = 20]             
node [ xshift=-3mm,yshift=0mm, align=center] 
{ $\neg T_2$} (S1)
    
(S1) edge [loop below]          
     node [xshift=10mm,yshift=5mm]   
     {True} (S1)
      
 (S3) edge [dashed, loop above]          
 node [xshift=10mm,yshift=-5mm]   
 {$T_1 \wedge T_2$} (S3)

(S0) edge [dashed, bend left  = 10] 
node [xshift=3mm,yshift=3mm,align=center] 
{$T_1 \wedge T_2$} (S3)
    
 (S2) edge [dashed, loop right] 
 node [xshift=-2mm,yshift=-4.5mm,align=center] 
 {$T_1 \wedge \neg T_2$} (S2)
     
(S3) edge [dashed, bend left = 10] 
node [xshift=0mm,yshift=0mm,align=center] 
{$T_1 \wedge \neg T_2$} (S2)

(S3) edge [dashed] 
node [xshift=0mm,yshift=0mm,align=center] 
{$\neg T_1$} (S1)

(S2) edge [dashed, bend left = 10]   node [xshift=0mm,yshift=0mm,align=center] 
{$\neg T_1 \vee T_2$} (S1)      
     
;
\end{tikzpicture}}
		\caption{Relaxation automaton $\mathcal{R}(I,O)$: \emph{impossible} means the system $\mathcal{D}$ will not allow the state to be reached, and the shield $\shield$ can treat it as \emph{don't care}.}
		\label{fig:relax}
	\end{minipage}
\vspace{2ex}

	\begin{minipage}[b]{\linewidth}
		\centering
		\scalebox{0.675}{\begin{tikzpicture}[auto,node distance=\nd]
\node[state,initial] at  (0,0)       (S0) {};
\node[state,accepting]         at  (4.5,0)     (S1) {infeasible};
\path
(S0) edge [loop above] 
     node [xshift=-15mm,yshift=-2mm,align=center] 
     {$(M'_1 \vee \neg M'_2)$  } (S0)

(S0) edge []             
     node [xshift=0mm,yshift=0mm, align=center] 
     {$(\neg M'_1 \wedge M'_2) $} (S1)
    
(S1) edge [loop above]          
     node [xshift=8mm,yshift=-5mm]   
     {True} (S1)
 
;
\end{tikzpicture}}
		\caption{Feasibility automaton $\mathcal{F}(O')$: \emph{infeasible} means the state is unrealizable, and the shield $\shield$ must avoid the related edges while generating solutions.}
		\label{fig:feasibility}
	\end{minipage}
\end{figure}

To check the compatibility of the predicate values, conceptually, one
can iterate through all possible value combinations for the predicates
in $\mathcal{P}$, and check each combination with an LP solver. If the
combination is \emph{unsatisfiable (UNSAT)} according to the LP
solver, we say it is impossible.  However, in our actual
implementation, the compatibility checking is performed significantly
more efficiently, due to the use of variable partitioning and UNSAT
cores.
First, $\mathcal{P}$ may be divided into subgroups, such that
predicates from different subgroups do not interfere with each
other. Therefore, value combinations may be computed via Cartesian
products.
Second, when a value combination is proved to be unsatisfiable, we
compute its UNSAT core, i.e., a minimal subset that itself is
UNSAT. By leveraging these UNSAT cores, we can significantly speed up
the checking of value combinations.

\ignore{

$\mathcal{R} = (R, r_0, \dalph, \delta^r, F^r)$ captures all the
impossible combinations of the corresponding predicates.

It consists of two parts. The first part, which is optional, is an automaton $\spec^c = (C, c_0, \dalph, \delta^c, F^v)$ provided by the users, who can leverage their understanding of the system to capture impossible
behaviors of $\mathcal{D}$. For example, $\spec^c$ may capture certain
aspect of a physical law, e.g., time can never travel backward; or certain properties that already satisfied 
by the system. 
The second part of the relaxation automaton is defined as an automaton $\spec^f = (F, f_0, \dalph_p, \delta^f, F^f)$, which is computed to capture infeasible combinations of the predicates $\dalph_p \in \dalph$ in $\omega$. 

\begin{itemize}
\item  $\delta^f(f, \dletter_p) = (\{f \in F \setminus F^f\})$, meaning that unsafe state $f$ is a trap state
\item  $\delta^f(f \in F^f, \dletter_p) = (\{f' \in F \setminus F^f\})$ if $\gamma^p$ is unsatisfiable
\item  $\delta^f(f \in F^f, \dletter_p) = (\{f' \in F^f\})$ if $\gamma^p$ is satisfiable
\end{itemize}

therefore, the relaxation automaton $\mathcal{R}$ is the synchronous product of $\spec^c$ and $\spec^f$, such that $R = C \times F$ is the state space, $r_0 = (c_0, f_0)$ is the initial state, $\delta^r$ is the transition function and $F^r$ is a set of safe state, such that $\delta^r((c,f), \dletter) = (\delta^c(c, \dletter), \delta^f((f, \dletter))$ and $F^r = \{(c,f) \in R \mid c \in F^c \wedge f \in F^f\}$

}

\subsection{Computing the Feasibility Automaton}

The feasibility automaton $\mathcal{F}$ aims to capture the
combinations of $O'$ values that are unrealizable in the real domain.
Similar to $\mathcal{R}$, states in $\mathcal{F}$ are divided into two
types: \emph{normal} and \emph{infeasible}.  Here, \emph{normal} means
the value combinations are realizable in the real domain, whereas
\emph{infeasible} means the value combinations may be unrealizable.

Fig.~\ref{fig:feasibility} shows an example feasibility automaton for
the running example: all predicates are the primed versions because
they are defined over $O'$ signals, which are part of the modified
output of the shield.
Upon $\neg M'_1 \wedge M'_2$, the automaton goes into the infeasible
state, because $\neg (|\mu'|<0.2) \wedge (|\mu'|<0.02)$ has no
real-valued solution.

While this is rare, a value combination may depend on real-valued
signals not only in the shield's output ($O'_r$) but also in the input
($I_r$).  Let such a value combination be denoted by $\phi(I_r,O'_r)$.
\emph{Whether $\phi$ is guaranteed to be realizable} can be decided
using an SMT solver, by checking the validity of the formula $\forall
I_r . \exists O'_r . \phi(I_r,O'_r)$.

Subsequently, during our computation of the winning strategy
$\omega_r$, we need to avoid such unrealizable combinations.

\ignore{
 $\mathcal{F}= (F, f_0, \dalph_p', \delta^f, F^f)$, where 
$\dalph_p' \in \doutalphprime$ in $\omega$. 
In most of the time, it is a projection of $\spec^f$ from $\doutalph$ variables to $\doutalphprime$.  
Only $\dalph_p'$ that cannot be mapped from $\dalph_p$ need to be recalculated.
For example, Fig.~\ref{fig:feasibility} shows the feasibility automaton of specifications in Section~\ref{sec:predicates}.
}

\subsection{Solving the Constrained Game}

The new safety game $\mathcal{G}_r$ is defined as the composition of
$\mathcal{W}$, the winning region of the Boolean game $\mathcal{G}$,
the relaxation automaton $\mathcal{R}$, and the feasibility automaton
$\mathcal{F}$.  We tweak the winning region automaton $\mathcal{W}$ by adding
an \emph{unsafe} state for all edges going out of $\mathcal{W}$.
Here, composition means the standard synchronous
product, where a state transition exists only if it is allowed by all
three components ($\mathcal{W}$, $\mathcal{R}$ and $\mathcal{F}$).
Furthermore, safe states of $\mathcal{G}_r$ are either (1) states that
are both \emph{safe} in $\mathcal{W}$ and \emph{feasible} in
$\mathcal{F}$, or (2) states that are \emph{impossible} in
$\mathcal{R}$.

More formally, assume that $F^\mathcal{W}$ is the set of unsafe states
related to the winning region $\mathcal{W}$, $F^\mathcal{F}$ is the
set of infeasible states of the feasibility automaton $\mathcal{F}$,
and $F^\mathcal{R}$ is the set of the impossible states of the
relaxation automaton $\mathcal{R}$.  The set of safe states in the new
game $\mathcal{G}_r$ is defined as $(\neg F^\mathcal{W} \wedge \neg
F^\mathcal{F}) \vee F^\mathcal{R}$.

Finally, we solve $\mathcal{G}_r$ using standard algorithms for safety
games, e.g., Mazala~\cite{Mazala01}, which are also used in the prior
work~\cite{BloemKKW15,WuZW16,KonighoferABHKT17}.
The result is a winning strategy $\omega_r$, which in turn may be
implemented as a reactive component $\shield_\mathit{bool}$.  Note
that $\shield_{bool}$ is a Mealy machine that takes $I$ and $O$
signals as input and returns the modified $O'$ signals as output.
Furthermore, due to the use of $\mathcal{R}$ and $\mathcal{F}$, the
output of $\shield_\mathit{bool}$ is guaranteed to be realizable at
run time.

\ignore{
Given the original winning strategy $\omega$,
relaxation automaton $\mathcal{R}$ and feasibility automaton
$\mathcal{F}$, we construct a new safety game
$\mathcal{G}_\mathit{feas} = (\gstates,
\ginit, \dalph, \doutalphprime, \delta, \fstates)$, which is the synchronous product of $\omega$, $\mathcal{R}$ and $\mathcal{F}$.
Specially, $\fstates = \{(w, r, f) \in \gstates \mid w \in W \wedge r \in F^r \wedge f \not\in F^f\}$.

By solving the safety game~\cite{Mazala01}
(cf.~\cite{BloemKKW15,WuZW16,KonighoferABHKT17}), we can compute the
winning strategy $\omega_\mathit{feas}$, which allows us to satisfy $\spec$
by choosing proper values of $O'$ that are always realizable if they are predicates over real-valued $O_r'$.
}

\section{Generating the Real-valued Signals}
\label{sec:real_shield}

In this section, we present our method for computing the real-valued
signals ($O'_r$) at run time, based on the Boolean shield's output
($O'$).

Algorithm~\ref{alg:shield} shows the details of our method, which
needs $I_r$, $O_r$, $O'_r$, the set $\mathcal{P}$ of predicates,
$\shield_\mathit{bool}$, and a utility function $\gamma$, which is
used to evaluate the quality of the real-valued solution.
First, real values in $I_r$ and $O_r$ are transformed to Boolean
values in $I$ and $O$. Then, they are used by $\shield_\mathit{bool}$
to compute new values in $O'$.  When $O'$ and $O$ have the same
Boolean value, meaning the shield does not make any correction, $O'_r$
and $O_r$ will also have the same real value; in this case, no
computation is needed (Line 5).  However, when $O'$ and $O$ have
different values, we need to recompute the real values in $O'_r$
(Lines 7-11).

\begin{algorithm}[t]
{\footnotesize
	\caption{\footnotesize Computing real-valued correction signals at run time.}
	\label{alg:shield}
	\begin{algorithmic}[1]
		\Function{computeRealValues}{ $I_r$, $O_r$, $O'_r$, $\mathcal{P}$, $\shield_\mathit{bool}$, $\gamma$}
		\State $I, O \gets$  \Call{genBooleanAbstraction}{$I_r, O_r, \mathcal{P}$}
		\State $O' \gets$  \Call{genBooleanShieldOutput}{$\shield_\mathit{bool}, I, O$}
		\If{$O' = O$}
		\State $O'_r = O_r$
		\Else{}
		\State \highlight{ $O'_r \gets$  \Call{prediction}{$Hist$} }
		\If{\highlight{ $\neg$ \Call{Satisfiable}{$\mathcal{P}, O', O'_r$}}}
		\State \highlight{$model \gets$\Call{lpSolve}{$\mathcal{P}, \gamma, O'$}}
		\State \highlight{$O'_r \gets model$ }
		\EndIf
		\EndIf
		\State \highlight{ $Hist \gets Hist \cup \{O'_r\}$ }
		\EndFunction
	\end{algorithmic}
}
\end{algorithm}

\subsection{Robustness Optimization}

Since the output of the Boolean shield is an assignment of the Boolean
predicates in $O'$, and each predicate corresponds to a linear
constraint of the form $\Sigma_{i=1}^k a_i x_i \leq 0$, conceptually,
the real values in $O'_r$ can be computed by solving the linear
programming (LP) problem.

However, naively invoking the LP solver does not always produce a
high-quality solution.  Instead, we develop the following optimization
to improve the quality of the solution.
Specifically, we restrict the LP problem using a robustness constraint
derived from the utility function $\gamma$.  While there may be
various ways of defining robustness, especially in the context of
STL~\cite{FainekosP06,DokhanchiHF14}, a straightforward way that works
in practice is to ensure the signal is \emph{smooth} (see
Fig.~\ref{fig:real_ex}).

That is, we restrict the LP problem using the objective function
\[
\min \Big(|val^i - \frac{\sum\limits_{k=1}^N val^{i-k}}{N}| \Big)
\]
where $val^i$ denotes the current value (at the $i$-th time step) and
$val^{i-k}$, where $k=1,2,\dots$, denotes the value in the recent
past.  The above function aims to minimize the distance between
$val^i$ and the (moving) average of the previous $N$ values, stored in
$\mathit{Hist}$ (Line~13).

\subsection{Value Prediction and Validation}

While the robustness constraint improves the quality of the
real-valued solution, it also increases the computational cost of LP
solving.  To reduce the computational cost, we develop a two-phase
optimization for computing the solution.

First, we predict the value of a signal using standard regression
algorithms based on the historical values of the signal in the immediate past
(Line 7 in Algorithm~\ref{alg:shield}).  Here, the procedure
$\proc{prediction}$ leverages historical values stored in $Hist$.
Since the signal is expected to be \emph{smooth}, standard linear or
non-linear regression can be very accurate in practice.

Next, we validate the predicted value (Line 8).  This is accomplished
by plugging the predicted value for $O'_r$ into the combination of
Boolean predicates defined by $\mathcal{P}$ and the values of signals
in $O'$.  If it is valid, the value is accepted as the final output,
and invocation of the LP solver is avoided.  Note that the time taken
to perform prediction and validation is significantly smaller than
that of the LP solving.

Only when the predicted value is not valid, we invoke the LP solver
(Line 9).  Even in this case, the response time is fast because we can
use the same LP solver for validation and LP solving.  Due to
incremental computation inside the solver, the solution used for
validation, which is often close to the final solution, can help speed
up LP solving.

%\newpage
\section{Experiments}
\label{sec:experiment}

We have implemented our method as a tool that takes the automaton
representation of a safety specification as input and returns a
real-valued shield as output.  Internally, we solve the safety game
using Mazala's algorithm~\cite{Mazala01} implemented symbolically
using CUDD~\cite{Somenz95}, and use the LP solver integrated in
Z3~\cite{de2008z3} for prediction, validation and constraint solving.
%
%As for real-valued signals, a set $\mathcal{P}$ of predicates is
%created to aid in the realizability analysis  and
%computation of real values at run time. 
%
For evaluation purposes, the shield is implemented as a C program and
is executed at every time step.  Each execution has two phases: (1)
generating Boolean values for signals in $O'$, and (2) generating real
values for signals in $O'_r$.

\vspace{1ex}
\noindent
\textbf{Benchmarks}~~
We evaluated our tool on seven sets of benchmarks, including automotive
powertrain control~\cite{JinHSCC2014}, autonomous
driving~\cite{raman2015reactive}, adaptive cruise
control~\cite{nilsson2016correct}, multi-drone fleet
control~\cite{pant2018fly}, generic control~\cite{jin2015mining},
blood glucose control~\cite{roohi2018parameter}, and water tank
control~\cite{AlshiekhBEKNT18}.
In all benchmarks, the original specification was given in STL, which
has both timing and real-valued constraints.

\begin{table}[t]
\centering
\caption{Statistics of the benchmark applications.}
\label{tab:bench}

\scalebox{0.7}{\begin{tabular}{|p{0.15\linewidth}|c|p{1.025\linewidth}|}
\hline
Application &Property   &STL Formula and Description \\
            &           &                            \\\hline\hline

	     &R26 & In normal mode, permitted overshoot/undershoot is always less than 0.05 \\
             &    & $\mathsf{G}_{[\tau_s,T]} \big( l\!=\!\textsf{normal} \Rightarrow |\mu| < 0.05 \big)$ \\
	\cline{2-3}
             &R27 & In normal mode, overshoot/undershoot less than 0.02 within the settling time \\
	     &    & $\mathsf{G}_{[\tau_s,T]} \Big( \textsf{rise}(a)|\textsf{fall}(a) \Rightarrow \mathsf{G}_{[\eta,\frac{\varsigma}{2}]} \big( |\mu| < 0.02 \big) \Big)$ \\ 
	\cline{2-3}
Powertrain   &R32 & From power to normal, overshoot/undershoot less than 0.02 within settling time\\
             &    & $\mathsf{G}_{[\tau_s,T]}  \Big( l\!=\!\textsf{power} \wedge \mathsf{X}(l\!=\!\textsf{normal}) \Rightarrow  \mathsf{G}_{[\eta,\frac{\varsigma}{2}]} \big( |\mu| < 0.02 \big) \Big)$ \\
	\cline{2-3}
	     &R33 & In power mode, permitted overshoot or undershoot should be less than 0.2 \\
             &    & $\mathsf{G}_{[\tau_s,T]} \big( l\!=\!\textsf{power} \Rightarrow |\mu| < 0.2 \big)$ \\
	\cline{2-3}
	     &R34 & Upon startup/sensor failure, overshoot/undershoot $<$0.1 within the settling time\\
	     &    & $\mathsf{G}_{[\tau_s,T]}
%			\begin{pmatrix}
			l\!=\!\textsf{startup}|\textsf{sensor\_fail} \wedge \textsf{rise}(a)|\textsf{fall}(a) \Rightarrow 
%\\
			\mathsf{G}_{[\eta,\frac{\varsigma}{2}]} \big( |\mu| < 0.1 \big)\Big)
%			\end{pmatrix}
			$ \\
	\hline

             &D1  &Vehicle should keep a steady speed $V_s$ when there is no collision risk \\
Autonomous   &    & $\mathsf{G} \big( |y^{ego}_k - x^{adv}_k| >= 4 \big) \Rightarrow \mathsf{G} \big( |v^{ego}_k - V_s| < \varepsilon \big) $ \\
	\cline{2-3}
Driving      &D2  &Vehicle should come to stop for at least 2 second when there is collision risk \\
                     &    &$\mathsf{G} \big( |y^{ego}_k - x^{adv}_k| < 4 \big) \Rightarrow \mathsf{G}_{[0,2]} \big( |v^{ego}_k| < 0.1 \big) $\\
	\hline

                        &A1 &Keep a safe distance with lead vehicle:~
                        $\mathsf{G} \big( \textsf{pos\_lead}[t] - \textsf{pos\_ego}[t] > D_s \big)$\\
	\cline{2-3}                
Cruise                 &A2 &Achieve cruise velocity if there is a comfortable distance  \\
Control                &   & $ \big( \textsf{pos\_lead}[t]  -     \textsf{pos\_ego}[t] > D_c \big) \mathsf{U}_{[0,10]}  \big( |\textsf{v\_ego}[t] - \textsf{v\_cruise}[t]| < \varepsilon \big)$\\
        \cline{2-3}
			&A3 &Vehicle should never travel backward:~
                        $\mathsf{G}\big( \textsf{v\_ego}[t] >= 0\big)$ \\
	\cline{2-3}
			&A4 &Vehicle doesn’t halt unless lead vehicle halts:\\
                       &   & $\mathsf{G}\big( \textsf{v\_lead}[t] > 0 \big) \Rightarrow \mathsf{G}\big( \textsf{v\_ego}[t] > 0 \big) $	\\
	\hline
                   &Q1   &Drone flies to goal point if no obstacles are on the way:\\
Quadrotor          &     &$\mathsf{G} \big( Obs(\mathbf{pos}^{quad}, \mathbf{pos}^{obs}) \Rightarrow \mathbf{\omega}_g >0 ) \big)$\\
	\cline{2-3}
 Control       	   &Q2   &Avoiding obstacles:$ \mathsf{G}  \neg Obs(\mathbf{pos}^{quad}, \mathbf{pos}^{obs}) \Rightarrow$\\
                   &     &
			$ 
			\Big( \mathbf{\omega}_{\bar{g}} >0  \wedge \mathsf{G} \big( Dis(\mathbf{pos}^{quad}, \mathbf{pos}^{obs}) < \varepsilon \Rightarrow   \mathbf{\omega}_g = 0 \big) \Big)$ \\
	\hline

                &C1   &After settling, output error should be less than set value $\varepsilon_b$:\\
General         &     &
                $\mathsf{G} \Big( \textsf{x}[t] \Rightarrow \mathsf{G}_{[10, \infty]}  \big(| \frac{  \textsf{y}[t] - \textsf{y}^{ref} } {\textsf{y}^{ref}} | <  \varepsilon_b\big)  \Big)$\\
	\cline{2-3}

Control         &C2   &Output error should be $[ \varepsilon^\bot,  \varepsilon^\top]$ in settling time:\\
                &     &
                $\mathsf{G} \Big( \textsf{x}[t] \Rightarrow \mathsf{G}_{[0, 20]}  \big( \varepsilon^\bot < \frac{  \textsf{y}[t] - \textsf{y}^{ref} } {\textsf{y}^{ref}}   <  \varepsilon^\top \big)  \Big)$\\
	\cline{2-3}
        	&C3   &Output should achieve reference value within $rise\_time$:\\
                &     &
                $\mathsf{G} \Big( \textsf{x}[t] \Rightarrow \mathsf{F}_{[0, \textsf{rise\_time}]}  \big(| \frac{  \textsf{y}[t] - \textsf{y}^{ref} } {\textsf{y}^{ref}} | <  \varepsilon_r\big)  \Big)$\\
	\hline

Glucose Control & B1 & Having meal within $t_1$ minutes after taking the bolus is safe. A bolus must be taken after $t_2$ minutes of having meal, if it is not yet taken:\\
                &    &
		      $\mathsf{G} \Big(    \mathsf{F}_{[0, t_1+t_2]} (B > \textsf{c}_2) \vee \mathsf{G}_{[t_1, t_1 + t_2]} \big( M > \textsf{c}_1 \Rightarrow  \mathsf{F}_{[0, t_2]} (B > \textsf{c}_2) \big)  \Big)$ \\
	\hline

Water Tank      &W1 &Turn on inflow and turn off outflow switch when water level is low ($l < 4$) \\
Control         &   & $\mathsf{G} \big( l < 4 \Rightarrow \mathsf{G}_{[0,3]} (\textsf{flow}_{out} =0 \wedge 1 < \textsf{flow}_{in} < 2) \big)$\\
	\cline{2-3}
		      &W2 &Turn on outflow and turn off inflow switch when water level is high ($l > 93$)  \\
                      &   &$\mathsf{G} \big( l > 93 \Rightarrow \mathsf{G}_{[0,3]} (\textsf{flow}_{in} =0 \wedge 0 < \textsf{flow}_{out} < 1) \big)$\\
	\hline
\end{tabular}}
\end{table}

Table~\ref{tab:bench} shows the benchmark statistics, including the
application name, the property, a short description, and the
corresponding STL formula. 
For brevity, we omit the 
automaton representations.
%, but they are available 
%online\footnote{Artifacts are available at \url{https://bitbucket.org/mengwu/shield-synthesis}}.
%
We conducted experiments on a computer with Intel i5 3.1GHz CPU, 8GB
RAM, and the Ubuntu 14.04 operating system.  Our experiments were
designed to answer the following questions: (1) Is our tool efficient
in synthesizing the real-valued shield?  (2) Is the shield effective
in preventing safety violations?  (3) Are the real-valued signals
produced by the shield of high quality?

\vspace{1ex}
\noindent
\textbf{Experimental Results}~
Table~\ref{tab:result} shows the results of our shield synthesis
procedure.  Columns~1-3 show the property name, the number of states
of the specification, and the number of real-valued signals in $I_r$
and $O_r$, respectively.
Column~4 shows the number of predicates defined over signals in $I_r$
and $O_r$.  Based on these predicates, Boolean signals in $I$ and $O$
are created; Column~5 shows the number of these signals.
Column~6 shows the number of conflicting constraints captured by the
relaxation and feasibility automata, respectively.
Column~7 shows the synthesis time.  Columns~8-9 show the
number of states of the Boolean shield, and the number of real-valued
constraints to be solved at run time.

\begin{table}%[h]
\centering
\caption{Results of our new shield synthesis procedure.}
\label{tab:result}
\scalebox{0.7}{\begin{tabular}{|p{0.2\linewidth}||c|c|c|c|c|r|r|c|}
\hline
     
Name & \multicolumn{2}{c|}{Specification }    & \multicolumn{4}{c|}{Synthesis Tool} & \multicolumn{2}{c|}{Shield $\shield$} \\ \cline{2-9}
     & states & $|I_r|$/$|O_r|$ &$|\mathcal{P}_I|$/$|\mathcal{P}_O|$ & $|I|$/$|O|$ & $|\mathcal{R}|$/$|\mathcal{F}|$ & time(s) & states & constrs\\ %&Game Size(\#State)
\hline\hline
R26+R27             &8  &1/1 &2/2 &5/2 &2/1  &0.16 &25   &2+2 \\\hline %640
R32+R33             &9  &1/1 &2/2 &5/2 &2/1  &0.15 &28   &2+2 \\\hline  %792
R26+R27+R32 +R33+R34 &23 &1/1 &2/4 &5/4 &12/11 &1.15 &158  &4+2 \\\hline  %6992 
D1                  &6  &3/1 &5/3 &6/3 &53/5  &0.15 &19   &3+2 \\\hline  %384
D2                  &5  &3/1 &2/3 &3/3 &5/5  &0.21 &30   &3+2 \\\hline  %400
D1+D2               &14 &3/1 &5/3 &6/3 &53/5  &0.8  &164  &3+2 \\\hline  %4032
A1+A3+A4            &3  &3/1 &2/2 &2/3 &1/1  &0.08 &8    &2+0 \\\hline  %96
A2+A3+A4            &4  &4/1 &3/3 &3/3 &4/4  &0.1  &15   &3+0 \\\hline   %192
A1+A2+A3+A4         &7  &4/1 &4/3 &4/4 &8/4  &0.55 &48   &3+0 \\\hline   %672
Q1+Q2               &5  &1/2 &1/2 &2/2 &0/0  &0.08 &7    &2+0 \\\hline   %280
C1+C2+C3            &19 &2/1 &3/4 &3/4 &13/11 &0.52 &118  &4+2 \\\hline   %4864
B1		            &5  &3/1 &5/1 &5/1 &14/0  &0.1  &6    &1+0 \\\hline   %200
W1+W2		        &6  &1/2 &2/2 &2/2 &1/0  &0.1  &10   &2+2 \\\hline   %384
\end{tabular}}
\end{table}

Table~\ref{tab:result2} shows the performance of the shields. For each
shield, we generated input signals (for $I_r$ and $O_r$) based on the
system description: some input signals satisfy the specification while
others do not.  By measuring the response time of the shield under
these input signals, and the quality of corrections made by the
shield, we hope to evaluate its effectiveness.

In this table, Column~1 shows the property name.  Column~2 shows the
size of the C program that implements the shield.  Column~3 shows the
response time of the Boolean shield on input signals that do not
violate the specification.  Columns~4-5 show the response time on
input signals that violate the specification.  Among these columns,
\emph{prediction} means the real-valued solution was successfully
computed by a linear regression, whereas \emph{constraint solving}
means prediction failed and the solution was computed by the LP
solver.

Overall, the time to compute real-valued correction signals is within
0.5 ms when $\mathcal{D}\not\models \spec$, and less than 1 us when
$\mathcal{D}\models \spec$.  In the latter case, the shield does not
need to make correction at all.  In both cases, the response time is
always bounded and fast enough for the target applications.

\begin{table}%[h]
\centering
\caption{Results of evaluating runtime performance of the shield.}
\label{tab:result2}
\scalebox{0.7}
{\begin{tabular}{|p{0.2\linewidth}|c|c|c|c|}
\hline

Name                & Implementation    &   \multicolumn{3}{c|}{Shield Response Time} \\\cline{3-5}
                    &    (LoC)          & Boolean step (us)     & prediction step (us)    &  constraint solving (us) \\\hline\hline
R26+R27             &745                &0.3      &293.3        &336.8\\\hline
R32+R33             &748                &0.41     &256.5        &333.9\\\hline
R26+R27+R32 +R33+R34 &1446               &0.8      &245.0        &279.8\\\hline
D1                  &781                &0.45     &177.2        &164.7\\\hline
D2                  &853                &0.5      &313.5        &329.0\\\hline
D1+D2               &2242               &0.8      &318.3        &202.4\\\hline
A1+A3+A4            &539                &0.37     &164.3        &212.7\\\hline
A2+A3+A4            &632                &0.49     &281.7        &431.5\\\hline
A1+A2+A3+A4         &940                &0.45     &291.7        &290.1\\\hline
Q1+Q2               &556                &0.18     &299.2        &313.5\\\hline
C1+C2+C3            &1037               &0.5      &299.4        &395.2\\\hline
B1                  &623                &0.31     &225.4        &313.4\\\hline
W1+W2               &608                &0.57     &295.3        &222.1\\\hline

\end{tabular}}
\end{table}

\begin{figure}
	\centering
	\includegraphics[width=.5\textwidth]{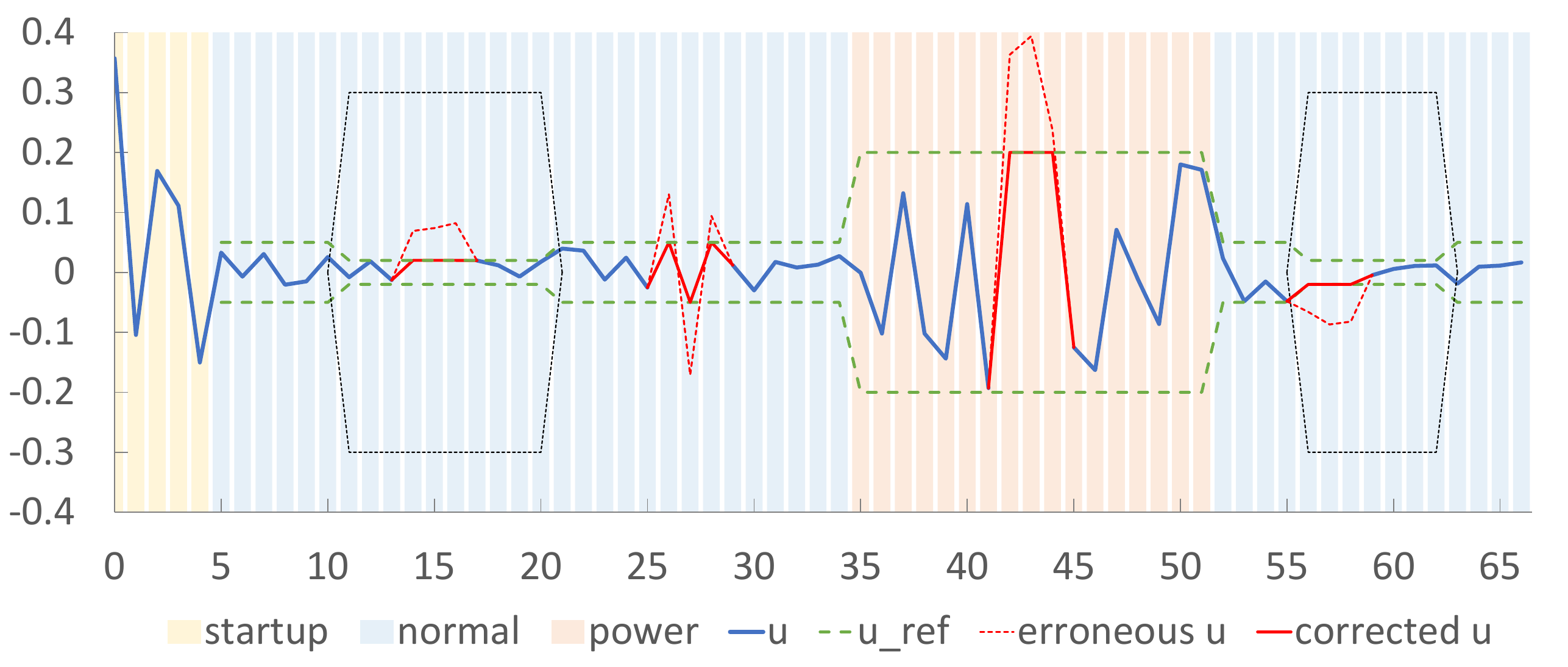}
	\caption{Automotive powertrain system simulation (w/ and w/o the shield).}
	\label{fig:toyotasim}
\end{figure}

\vspace{1ex}
\noindent
\textbf{Case Study 1: Powertrain Control System}~~
To validate the effectiveness of our approach, we integrated the
shield into the simulation model of the powertrain control system.
Then, we compared the system performance with and without the shield.
Fig.~\ref{fig:toyotasim} shows the simulation results, where our
shield was synthesized from the system properties 26, 27, 32, 33 and
34 as described in Jin et al.~\cite{JinHSCC2014}. Recall that $\mu$ is
the normalized error of the A/F ratio and $\mu_\mathit{ref}$ is a
reference value.

%Although the error signal is a real value, it is not a
%continues number as it is defined regarding the turbulence on the
%reference air-fuel ratio. Prediction is less usefully in such a case,
%but we define the robustness function as $\Max |\mu|$ to allow maximum
%relaxation for the system as long as it is safe.  

The green dashed line indicates the safe region, which varies as
the system switches between different modes (transition events are
highlighted with black dotted line). The red dashed line represents
violations of the specification by the $O_r$ signals.  The
solid red line represents corrections made in $O'_r$.
The result shows that our shield can always produce real-valued
correction to keep $\mu$ in the safe region.

Recall that the goal of using a shield is not to correct the flawed
design $\mathcal{D}$ itself, which includes the overshot \emph{plant};
instead, the goal is to avoid the negative impact of $\mathcal{D}$'s
output.  Here, the output signal $\mu$ may be used by other components
of the system.

\begin{figure}

	\begin{minipage}{\linewidth}
	\centering
		\includegraphics[width=\textwidth]{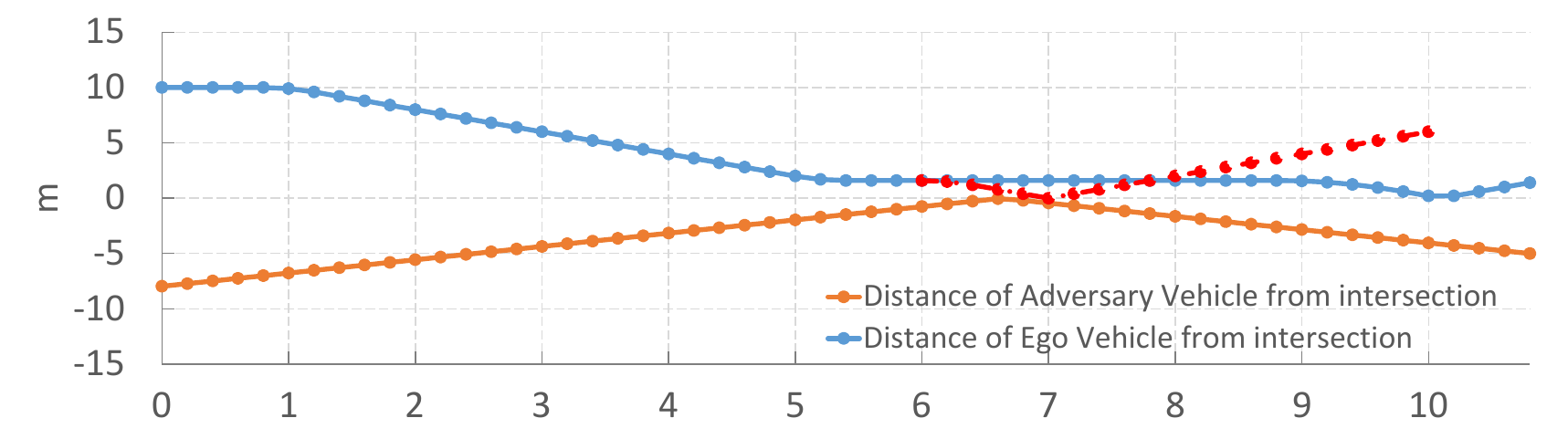}
	\end{minipage}

\vspace{1ex}
	\begin{minipage}{\linewidth}
	\centering
			\includegraphics[width=\textwidth]{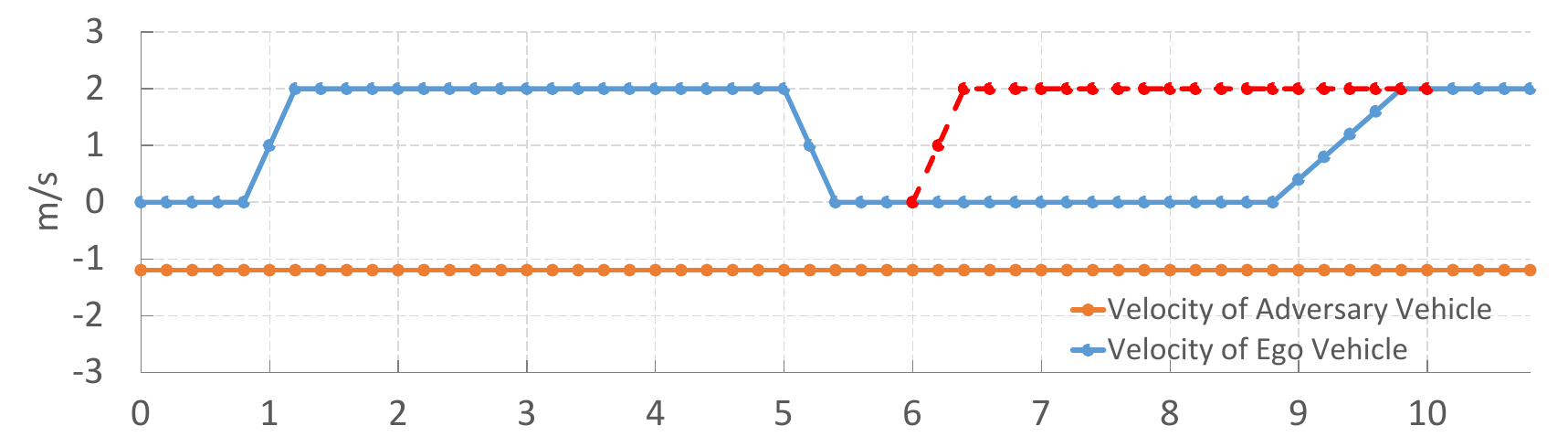}
	\end{minipage}

\vspace{1ex}
	\begin{minipage}{\linewidth}
	\centering
			\includegraphics[width=\textwidth]{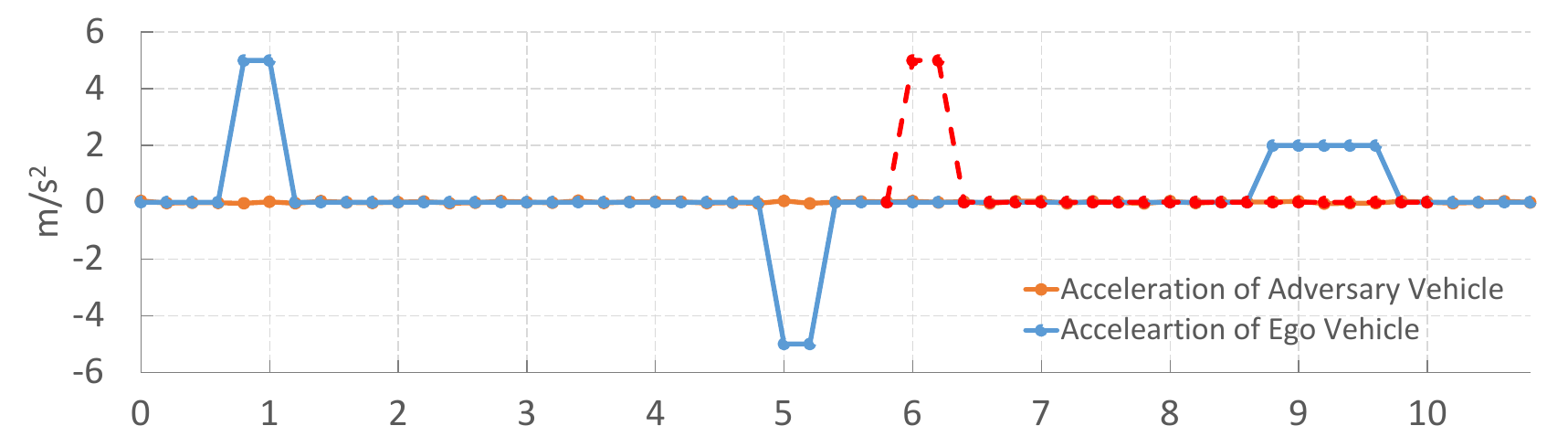}
	\end{minipage}

\vspace{1ex}
	\caption{Position, velocity and acceleration in autonomous driving simulation.} \label{fig:autodrive_sim}
\end{figure}

\vspace{1ex}
\noindent
\textbf{Case Study 2: Autonomous Driving}~~ 
Fig.~\ref{fig:autodrive_sim} shows the simulation results of an
autonomous driving system~\cite{raman2015reactive} with and without
our shield.  Here, an ego vehicle is put into a nondeterministic
environment that includes an adversarial vehicle, and the two cars are
crossing an intersection.  The ego vehicle is protected by a shield
synthesized from D1+D2 in Table~\ref{tab:bench}.
The three plots, from top to bottom, are for distances to the
intersection, velocities, and accelerations of the two vehicles.  The
$x$-axis represents the time in seconds.

The adversarial vehicle drives straight through the intersection at a
constant speed.  The ego vehicle, in contrast, may change speed to
avoid collision.  From $t=0s$ to $t=5s$, since the distance between
the two vehicles is large, the ego vehicle maintains a steady speed (set to
$2m/s$ initially). At $t=5s$, based on the safety specification, it is
supposed to come to a stop (for at least $2s$ or when there is no
collision risk).  However, since we injected an error at $t=6s$ (in
red dashed line), there is an unexpected acceleration and, without the
shield, there would have been  a collision.

The blue lines show the behavior of the ego vehicle after corrections
are made by the shield.  Clearly, its behavior satisfies the
requirements: it stops at the intersection to allow the adversarial
vehicle to pass safely. Furthermore, the real-valued correction made
by the shield is successfully predicted using linear regression, and
the predicted values satisfy not only the safety but also the
robustness requirements.

%\newpage
\section{Related Work}
\label{sec:related}

%We are the first to synthesize real-valued shields and demonstrate
%their application to cyber-physical systems.  
As we have mentioned earlier, prior work on shield synthesis has been
restricted to the Boolean domain.  Specifically, Bloem et
al.~\cite{BloemKKW15} introduced the notion of shield together with a
synthesis algorithm, which minimizes the deviation between $O$ and
$O'$ under the assumption that \emph{no two errors occur within $k$
  steps}.
%If there are more errors within $k$ steps, the shield would
%stop minimizing deviation although it still ensures safety of the
%composed system.
%
Wu et al.~\cite{WuZW16} improved the algorithm to deal
with \emph{burst error}.  That is, if more errors occur within the
$k$-step recovery period, instead of entering a \emph{fail-safe}
state, they keep minimizing the deviation.  K{\"{o}}nighofer et
al.~\cite{KonighoferABHKT17} further improved the shield while
Alshiekh et al.~\cite{AlshiekhBEKNT18} leveraged it to improve the
performance of reinforcement learning.
However, none of the existing techniques dealt with the realizability
problems associated with real-valued signals.

%In addition, these previous work
%mainly focus on synthesizing a 
%controller and solving optimization problems for outputting the most 
%satisfied control sequences. As opposed to that, our work pays more attention
%to synthesizing the shield that correcting the controller's output during runtime.
%Due to the safety guarantee, our approach can substantially increase the robustness 
%of a controller with unexpected disturbance while
%~\cite{raman2014model,liu2011synthesis,farahani2015robust,raman2015reactive} may
%fail to return back to normal status.

There is also a large body of work on reactive
synthesis~\cite{Pnueli89,BloemJPPS12,SohailS13,EhlersT14} and
controller
synthesis~\cite{raman2014model,liu2011synthesis,farahani2015robust,raman2015reactive}.
The goal is to synthesize $\mathcal{D}$ from a complete specification
$\Psi$, or the control sequences for $\mathcal{D}$ to satisfy $\Psi$.
In both cases, the complexity depends on $\mathcal{D}$.
This is a more challenging problem, for two reasons.
First, specifying all aspects of the system requirement may be
difficult.  Second, even if $\Psi$ is available, synthesizing
$\mathcal{D}$ from $\Psi$ is difficult due to the
inherent \emph{double exponential} complexity of the synthesis
problem.  
Our method, in contrast, treats $\mathcal{D}$ as a blackbox while
focusing on a small subset $\spec \subseteq \Psi$
of \emph{safety-critical} properties.  This is why shield synthesis
may succeed where conventional reactive synthesis fails.

Renard et al.~\cite{RenardFRPJM15} proposed a runtime enforcement
method for timed automata, but assumed that controllable input events
may be delayed or suppressed, whereas our method does not require such
an assumption.  Bauer et al.~\cite{Bauer2011} and Falcone et
al.~\cite{FalconeFM12} studied various types of temporal logic
properties that may be monitored or enforced at run time.  Renard et
al.~\cite{RenardRF17} also leveraged B\"uchi games to enforce regular
properties with uncontrollable events.  Our work is orthogonal in that
it tackles the realizability and efficiency problems associated with
real-valued signals.  Furthermore, we focus on safety while leaving
liveness properties and
hyper-properties~\cite{BonakdarpourF18,FarzanV19,CoenenFST19} for
future work.

An important feature of the shield synthesized by our method is that
it always makes corrections \emph{instantaneously}, without any delay.
Therefore, it differs from a variety of solutions that allow delayed
corrections.  In some cases, for example, buffers may be allowed to
store the erroneous output temporarily, before computing the
corrections~\cite{Schneider00,LigattiBW09,FalconeFM12}.  In this
context, the notion of \emph{edit-distance} is more relevant.  Yu et
al.~\cite{yu2011}, for example, proposed a technique for minimizing
the edit-distance between two strings, but the technique requires the
entire input be stored in a buffer prior to generating the output.
However, when the buffer size reduces to zero, these existing
techniques would no longer work.

Runtime enforcement is related to, but also different from, the
various software techniques for error avoidance.  For example,
failure-oblivious computing~\cite{RinardCDRLB04,LongSR14} was used to
allow software applications to execute through memory errors; temporal
properties~\cite{luo2013,ZhangW14} were leveraged to control thread
schedules to avoid runtime failures of concurrent software.  However,
these techniques are not designed to target cyber-physical systems
with real-valued signals, where corrections are expected to be made
instantaneously, i.e., in the same time step when errors occur.

\section{Conclusions}
\label{sec:conclusion}

We have presented a method for synthesizing real-valued shields to
enforce the safety of cyber-physical systems.  The method relies on a
principled technique at the synthesis time to rule out impossible and
infeasible scenarios and ensure the realizability of real-valued
corrections at run time. We also proposed optimizations to speed up
the computation and improve the quality of the solution.  We have
evaluated our method on a number of applications, including case
studies with an automotive powertrain control system and an autonomous
driving system.  Our results demonstrate the effectiveness of the
method in enforcing safety properties of cyber-physical systems.

\section*{Acknowledgment}

This work was supported by the U.S.\ National Science Foundation (NSF) under
grants CNS-1813117, CNS-1722710, and CCF-1837131.
%
%Any opinions, findings, and conclusions expressed in this material are
%those of the authors and do not necessarily reflect the views of the
%funding agencies.

\newpage 
\bibliographystyle{plain}
\bibliography{shield}

\end{document}